\documentclass[aps,prd,twocolumn,groupedaddress]{revtex4-2}

\usepackage[utf8]{inputenc}
\usepackage[english]{babel}

\usepackage[dvipsnames]{xcolor}
\usepackage{amsmath,amssymb,enumerate,float,graphicx,hyperref,mathtools,slashed,pdfpages,empheq,tabularx,caption,multirow}

\allowdisplaybreaks

\graphicspath{{./figures/}}

\usepackage[most]{tcolorbox}

\tcbset{
	frame code={}
	center title,
	left=0pt,
	right=0pt,
	top=0pt,
	bottom=0pt,
	colback=yellow!25,
	colframe=white,
	width=\dimexpr\textwidth\relax,
	enlarge left by=0mm,
	boxsep=5pt,
	arc=0pt,outer arc=0pt,
}

\newcommand\myshade{85}
\hypersetup{
	linkcolor  = violet!\myshade!black,
	citecolor  = YellowOrange!\myshade!black,
	urlcolor   = Aquamarine!\myshade!black,
	colorlinks = true,
}

\newcommand{\pp}[1]{\left ( #1 \right )}
\newcommand{\bb}[1]{\left [ #1 \right ]}
\newcommand{\cc}[1]{\left \{ #1 \right \}}

\newcommand{\vep}{\varepsilon}

\newcommand{\bramket}[2]{\left\langle #1 \mid  #2 \right\rangle}
\newcommand{\bracket}[3]{\left\langle #1 \left\vert #2 \right\vert #3 \right\rangle}

\newcommand*\dif{\mathop{}\!\mathrm{d}}

\newcommand{\e}[1]{\mathrm{e}^{#1}}

\newcommand{\nn}{\nonumber\\ &}
\newcommand{\nneq}{\nonumber\\ & \hphantom{=}}
\newcommand{\nnhp}[1]{\nonumber\\ & \hphantom{#1}}

\makeatletter
\newcommand{\pushright}[1]{\ifmeasuring@#1\else\omit\hfill$\displaystyle#1$\fi\ignorespaces}
\newcommand{\pushleft}[1]{\ifmeasuring@#1\else\omit$\displaystyle#1$\hfill\fi\ignorespaces}
\makeatother

\makeatletter
\AtBeginDocument{\let\LS@rot\@undefined}
\makeatother

\begin{document}


\title{Determination of the strong vertices of doubly heavy baryons with pseudoscalar mesons in QCD}


\author{H. I. Alrebdi}
\email[]{hialrebdi@pnu.edu.sa}
\affiliation{Department of Physics, Princess Nourah bint Abdulrahman University, P.O. Box 84428, Riyadh 11671, Saudi Arabia}

\author{T. M. Aliev}
\email[]{taliev@metu.edu.tr}
\affiliation{Physics Department, Middle East Technical University, Ankara 06800, Turkey}

\author{K. \c Sim\c sek}
\email[]{ksimsek@u.northwestern.edu}
\affiliation{Department of Physics \& Astronomy, Northwestern University, Evanston, Illinois 60208, USA}


\date{\today}

\begin{abstract}
The strong coupling constant of doubly heavy baryons with light pseudoscalar mesons $ \pi $ and $ K $ are computed within the light cone sum rules. We take into account  two-particle and three-particle distribution amplitudes of the said pseudoscalar mesons. We compare our result with the one existing in the literature.
\end{abstract}


\maketitle

\section{Introduction}
The quark model has been very successful in studying the spectroscopy of baryons \cite{ref:1}. Many states of baryons predicted by the quark model have already been observed in experiments. For instance, practically all baryons containing single heavy quark have been observed in experiments.
\par 
The quark model also predicted the existence of the baryon family composed of two heavy and one light quarks. During the last two decades, many experimental efforts have been made for observation of these states \cite{ref:2, ref:3, ref:4}. The first experimental evidence of the doubly heavy baryon $ \Xi _{cc} $ with mass 3520 MeV in the channels $ \Xi _{cc} ^+ \to \Lambda _c^+ K^- \pi^+ $ and $ \Xi _{cc}^+ \to pD^+K^- $ was found by the SELEX Collaboration. Three years ago, the LHCb Collaboration announced the observation of $ \Xi _{cc}^{++} $ through the process $ \Xi _{cc}^{++} \to \Lambda _c^+ K ^- \pi ^+ \pi ^+ $ with mass $ (3624.40 \pm 0.72 \pm 0.14) $ MeV \cite{ref:5}. Later, the LHCb Collaboration measured the lifetime of $ \Xi _{cc}^{++} $ and confirmed the existence of $ \Xi _{cc}^{++} $ in the decay channel $ \Xi _{cc}^+ \pi ^+ $ \cite{ref:6}. The search of other doubly heavy baryons predicted by the quark model is now one of the main research areas in collider experiments \cite{ref:7, ref:8}. These observations stimulated a lot of theoretical studies, which can shed light on a deeper understanding of the inner structure of these baryons.
\par 
The study of the spectroscopy of doubly heavy baryons has been at the heart of tremendous theoretical studies. Within the framework of the Hamilton method \cite{ref:9}, the hypercentral method \cite{ref:10}, the lattice QCD \cite{ref:11, ref:12}, the QCD sum rules \cite{ref:13, ref:14, ref:15, ref:16, ref:17, ref:18}, the Bethe-Salpeter equation \cite{ref:19}, and in an extended chromomagnetic model \cite{ref:20}, the spectroscopy of doubly heavy baryons has been completely studied. 
\par 
For a deeper understanding of the dynamics of doubly heavy baryons, the study of their weak decays and strong and electromagnetic transitions is an ideal place. 
\par 
The semileptonic decays of doubly heavy baryons are analyzed within the QCD sum rules \cite{ref:21}, and the transitions $ \Xi _{QQ'} \to \Lambda _{Q'} $ and $ \Xi _{QQ'} \to \Sigma _{Q'} $ are studied within the light cone sum rules (\cite{ref:22} and \cite{ref:23}, respectively), in the framework of the light front formalism \cite{ref:24}, in the nonrelativistic quark model \cite{ref:25}, in the relativistic quark model \cite{ref:26}, and in the covariant constituent quark model \cite{ref:27}. However, the main studies focused on spectroscopic properties and weak decays of doubly heavy baryons. The studies of the strong and electromagnetic decays of doubly heavy baryons are very limited. Therefore, the study of their strong transitions would be timely. 
\par 
In the present work, we study the strong coupling constants of doubly heavy baryons with the light pseudoscalar mesons $ \pi $ and $ K $ within the light cone sum rules (LCSR) (For a discussion on the LCSR method, see for example \cite{ref:28}). Note that the strong coupling constants $ \Xi _{cc} \Xi _{cc} \pi $ and $ \Xi _{bb} \Xi _{bb} \pi $ within the same framework are studied in \cite{ref:29}. 
\par 
The paper is organized as follows. In Sec. \ref{sec:2}, we derive the LCSR for the strong coupling constants of doubly heavy baryons with pseudoscalar mesons $ \pi $ and $ K $. In this section, we present the details of the calculations for the strong coupling constants. Section \ref{sec:3} is devoted to the numerical analysis of the sum rules for the strong coupling constants. This section also contains our summary. 

\section{LCSR for the strong coupling constants of doubly heavy baryons with pseudoscalar mesons}\label{sec:2}
In order to determine the strong coupling constants of doubly heavy baryons with pseudoscalar mesons within the LCSR, we introduce the following correlation function:
\begin{align}
\Pi = i \int \dif ^4 x \ \e{ipx} \bracket{\mathcal P (q)}{\eta (x) \bar \eta (0)}{0} \label{eq:1}
\end{align}
where $ \mathcal P (q) $ is a pseudoscalar meson with momentum $ q $ and $ \eta $ denotes the interpolating current of the corresponding doubly heavy baryon. The $ SU(3) $ classification leads to the fact that there are two types of currents: symmetric and antisymmetric with respect to the exchange of two heavy quarks. The antisymmetric current exists only in the case in which two heavy quarks are different. The general forms of the interpolating currents (symmetric and antisymmetric) for doubly heavy baryons with $ J=1/2 $ can be written as 
\begin{align}
\eta ^{(S)} = \frac{1}{\sqrt 2} \vep ^{abc} \sum _{i=1}^2 \cc{
	\bb{ {Q^a}^{\rm T} A _1^i q ^b } A _2^i {Q'}^c 
	+ (Q \leftrightarrow Q')
} \label{eq:2}
\end{align}
and
\begin{align}
\eta ^{(A)} =& \frac{1}{\sqrt 6} \vep ^{abc} \sum _{i=1}^2 \Big\{
	2 \pp{{Q ^a}^{\rm T} A _1^i {Q'}^b } A _2^i q ^c \nn 
	+ \pp{{Q^a}^{\rm T} A _1^i q ^b} A _2^i {Q'}^c 
	- \pp{{{Q'}^a}^{\rm T} A _1^i q ^b} A _2^i Q ^C
\Big\} \label{eq:3}
\end{align}
where $ a $, $ b $, and $ c $ are color indices
, and
\begin{align}
A _1^1 = C, \quad A _1^2 = C \gamma _5,\quad A _2^1 = \gamma _5,\quad A _2^2 = \beta I \label{eq:4}
\end{align}
where $ \beta $ is an arbitrary parameter and $ C $ is the charge conjugation operator.
\par 
The main idea of the LCSR is the calculation of the correlation function in two different domains. On one hand, the correlation function is calculated in terms of hadrons. On the other hand, it is calculated in the deep Euclidean domain, $ p^2 \ll 0 $ and $ (p+q)^2 \ll 0 $, by using the operator product expansion (OPE) over twist. Then, performing the corresponding Borel transformation in order to suppress the contributions from higher states and the continuum and to enhance the contributions of the ground state, and matching these results, we can get the desired sum rules. 
\par 
The representation of the correlation function in terms of hadrons is obtained by inserting a complete set of baryon states carrying the same quantum numbers as the interpolating currents and by isolating the contribution of the corresponding ground states, namely
\begin{align}
\Pi =& \frac{\bracket{0}{\eta}{B _2(p _2)} \bramket{\mathcal P B (p _2)}{B (p _1)} \bracket{B(p _1)}{\bar \eta}{0}}{(p _2^2 - m _{B _2}^2) (p _1^2 - m _{B _1}^2)} \nn + \mbox{higher states} \label{eq:5}
\end{align}
where $ m _{B _2} $ and $ m _{B _1} $ are the masses of the final and initial doubly heavy baryons, respectively. The matrix elements in Eq. \eqref{eq:5} are determined as
\begin{align}
\begin{split}
\bracket{0}{\eta}{B _2 (p _2)} &= \lambda _{B _2} u (p _2)\\
\bramket{\mathcal P B _2(p _2)}{B _1 (p _1)} &= g _{B _1 B _2 \mathcal P} \bar u (p _2) i \gamma _5 u (p _1)
\end{split} \label{eq:6}
\end{align}
where $ \lambda $ is the residue, and $ g _{B _1 B _2 \mathcal P} $ is the relevant coupling constant of the doubly heavy baryons with the corresponding pseudoscalar meson. Taking into account \eqref{eq:6} in \eqref{eq:5} and performing the summation over Dirac bispinors for the physical part of the correlation function, we get
\begin{align}
\Pi = \frac{\lambda _{B _1} \lambda _{B _2} (\slashed p + m _{B _2}) [i \gamma _5] (\slashed p + \slashed q + m _{B _1}) g _{B _1 B _2 \mathcal P}}{(p^2 - m _{B _2}^2) [(p+q)^2 - m _{B _1}^2]} + \cdots \label{eq:7}
\end{align}
where we denote $ p = p _2 $ and $ p _1 = p + q $. Among all possible structures, we choose the structure $ \slashed p \slashed q \gamma _5 $ which contains the maximal number of external momenta, which usually leads to a more reliable stability, namely better predictions for the physical quantities. As a result for the physical part of the correlation function (i.e focusing on the structure $ \slashed p \slashed q \gamma _5 $), we get
\begin{align}
\Pi = - \frac{\lambda _{B _1} \lambda _{B _2} g _{B _1B _2\mathcal P}}{(p^2 - m _{B _2}^2) [(p + q)^2 - m _{B _1}^2]} \label{eq:8}
\end{align}
Performing a Borel transformation over variables $ -p^2 $ and $ -(p+q)^2 $, we obtain
\begin{align}
\Pi ^{(B)} = \lambda _{B _1} \lambda _{B _2} g _{B _1B _2\mathcal P} \e{-(m _{B _1}^2 + m _{B _2}^2)/2M^2} \label{eq:9}
\end{align}
On the other hand, the correlation function is calculated from the QCD side by using the OPE over twist. It involves the heavy quark propagator in the presence of a background field as follows from \eqref{eq:1} after applying Wick's theorem. As a result, we get
	\begin{align}
	\Pi ^{(SS)} &=  \frac 12 \vep ^{abc} \vep ^{a'b'c'} \int \dif^4 x \ \e{ipx} \sum _{ij} (A _1^i) _{\alpha\beta} (A _2^i) _{\rho\gamma}
	 (\tilde A _2^j) _{\gamma'\rho'}
	 \nneq \times (\tilde A _1^j) _{\alpha'\beta' } \langle \mathcal P (q) \vert \{
	 [S _{Q'\gamma\gamma'} ^{cc'} S _{Q\alpha\beta'}^{aa'} + (Q\leftrightarrow Q')
	 \nneq - S _{Q\alpha\gamma'}^{ac'} S _{Q'\gamma\beta'}^{ca'} - (Q\leftrightarrow Q')] q _\beta ^b \bar q _{\alpha'} ^{b'}
	 \} \vert 0 \rangle  \label{eq:10} \\
	\Pi ^{(AA)} &= \frac 16 \vep ^{abc} \vep ^{a'b'c'} \int \dif ^4 x \ \e{ipx} \sum _{ij} (A _1^i) _{\alpha\beta} (A _2^i) _{\rho\gamma} (\tilde A _2^j) _{\gamma'\rho'}
	\nneq \times (\tilde A _1^j) _{\alpha'\beta'} \langle \mathcal P (q) \vert \{ 4 S _{Q'\beta\alpha'} ^{bb'} S _{Q\alpha\beta'} ^{aa'} q _\gamma ^c \bar q _{\gamma'} ^c 
	\nneq - 2 S _{Q'\beta \gamma'} ^{bc'} S _{Q\alpha\beta'} ^{aa'} q _\gamma ^c \bar q _{\alpha'}^{b'} - 2 S _{Q\alpha\gamma'} ^{ac'} S _{Q'\beta\beta'} ^{ba'} q _{\gamma}^c \bar q _{\alpha'} ^{b'} 
	\nneq - 2 S _{Q'\gamma \alpha'} ^{cb'} S _{Q\alpha\beta'} ^{aa'} q _\beta ^b \bar q _{\gamma'} ^{c'} + S _{Q'\gamma\gamma'} ^{cc'} S _{Q\alpha\beta'} ^{aa'} q _\beta \bar q _{\alpha'}^b 
	\nneq + q _\beta ^b \bar q _{\alpha'} ^{b'} S _{Q\alpha\gamma'} ^{ac'} S _{Q'\gamma\beta'} ^{ca'} - 2 q _\beta ^b \bar q _{\gamma '} ^{c'} S _{Q'\alpha\alpha'} ^{ab'} S _{Q\gamma\beta'} ^{ca'}
	\nneq + q _\beta ^b \bar q _{\alpha'} ^{b'} S _{Q'\alpha\gamma'} ^{ac'} S _{Q\gamma\beta'} ^{ca'}  + q _\beta ^b \bar q _{\alpha'} ^{b'} S _{Q\gamma\gamma'} ^{cc'} S _{Q'\alpha\beta'} ^{aa'} \} \vert 0 \rangle \label{eq:11}\\
	\Pi ^{(SA)} &= \frac{1}{\sqrt {12}}  \vep ^{abc} \vep ^{a'b'c'} \int \dif^4 x \ \e{ipx} \sum _{ij} (A _1^i) _{\alpha\beta} (A _2^i) _{\rho\gamma}
	\nneq \times (\tilde A _2^j) _{\gamma'\rho'} (\tilde A _1^j) _{\alpha'\beta'} \langle \mathcal P (q) \vert \{ -2 q _\gamma ^c \bar q _{\alpha'} ^{b'} S _{Q'\beta\gamma'} ^{bc'} S _{Q\alpha\beta'} ^{aa'}
	\nneq  + 2 q _\gamma ^c \bar q _{\alpha'} ^{b'} S _{Q\alpha\gamma'}^{ac'} S _{Q'\beta\beta'} ^{ba'}  - q _\beta ^b \bar q _{\alpha'}^{b'} S _{Q'\gamma\gamma'} ^{cc'} S _{Q\alpha\beta'} ^{aa'} 
	\nneq + q _\beta ^b \bar q _{\alpha'} ^{b'} S _{Q\alpha\gamma'} ^{ab'} S _{Q'\gamma\beta'} ^{ca'}  + q _\beta ^b \bar q _{\alpha'} ^{b'} S _{Q'\alpha\gamma'} ^{ac} S _{Q\gamma\beta'} ^{ca'} 
	\nneq - q _\beta ^b \bar q _{\alpha'} ^{b'} S _{Q'\alpha\beta'} ^{aa'} S _{Q\gamma\gamma'} ^{cc'} \} \vert 0 \rangle  \label{eq:12}	
	\end{align}
In these expressions, the superscripts $ (SS) $, $ (AA) $, and $ (SA) $ denote the symmetry property of the current $ \eta $ and $ \bar \eta $, and $ \tilde A  _i = \gamma ^0 A _i^\dagger \gamma ^0 $. The heavy quark propagator in the presence of a background field in the coordinate space is
\begin{align}
S _{Q\alpha\beta} ^{aa'} &= \frac{m _Q^2}{4\pi^2} \bb{
	\frac{i K _2(m _Q\sqrt{-x^2}) \slashed x}{(\sqrt{-x^2})^2} + \frac{K _1(m _Q\sqrt{-x^2})}{\sqrt{-x^2}} 
}_{\alpha\beta} \delta ^{aa'} \nneq - \frac{g _s}{16\pi^2} m _Q \int _0^1 \dif u \Big\{
	\frac{i K _1(m _Q\sqrt{-x^2})}{\sqrt{-x^2}} [\bar u \slashed x \sigma _{\lambda\tau}
	\nneq + u \sigma _{\lambda \tau} \slashed x ] + K _0(m _Q\sqrt{-x^2}) \sigma _{\lambda \tau}		
\Big\} _{\alpha\beta} G _{\lambda \tau} ^{(n)} \pp{\frac{\lambda ^n}{2}} ^{aa'} \label{eq:13}
\end{align}
where $ G _{\lambda \tau} ^{(n)} $ is the gluon field strength tensor, the $ \lambda ^n $ are the Gell-Mann matrices, and the $ K _i (m _Q \sqrt{-x^2}) $ are the modified Bessel functions of the second kind. Now, using the Fiertz identities, 
\begin{align}
q _\alpha ^b \bar q _\beta ^{b'} \to - \frac{1}{12} (\Gamma _i) _{\alpha\beta} \delta ^{bb'} \bar q \Gamma _i q \label{eq:14}
\end{align}
and
\begin{align}
q _\alpha ^b \bar q _\beta ^{b'} G _{\lambda \tau} ^{(n)} \to -\frac 14 \frac 14 \pp{\frac{\lambda ^n}{2}} ^{bb'} (\Gamma _i) _{\alpha\beta} \bar q \Gamma _i G _{\lambda \tau} ^{(n)} q \label{eq:15}
\end{align}
we see that the following matrix elements appear in the calculation:
\begin{align}
\bracket{\mathcal P (q)}{\bar q \Gamma _i q}{0} \quad \mbox{and} \quad \bracket{\mathcal P (q)}{\bar q \Gamma _i G _{\lambda \tau} ^{(n)} q}{0} \label{eq:16}
\end{align}
In the expressions above, $ \Gamma _1 = I $, $ \Gamma _2 = \gamma _5 $, $ \Gamma _3 = \gamma _\alpha $, $ \Gamma _4 = i\gamma _\alpha\gamma _5 $, and $ \Gamma _5 = \frac{1}{\sqrt 2} \sigma _{\alpha\beta} $. These matrix elements are defined in terms of pseudoscalar meson distribution amplitudes (DAs), whose expressions are presented in Appendix \hyperref[app:A]{A}.
\par 
Inserting Eqs. \eqref{eq:13}--\eqref{eq:15} into Eqs. \eqref{eq:10}--\eqref{eq:12}, performing necessary calculations for the theoretical part of the correlation function, and doing the doubly Borel transformation over variables $ -p^2 $ and $ -(p+q)^2 $, we get the following results:
\begin{widetext}
	\begin{align}
	\Pi ^{(SS){\rm theo}} &= \frac{M^2 }{576 \pi ^2 m _{Q'}} \{
		9 M^2 (1-\beta)^2 m _Q \mu _{\mathcal P} \mathcal I _1 ^{12} ((1-2u)\mathcal T (\alpha _i)) \nneq + M^2 (-1+\beta) m _{Q'} [
			(-1+\beta) \mu _{\mathcal P} [9 \mathcal I _1^{12} ((1-2u)\mathcal T (\alpha _i))  - 56 (-1+\tilde \mu _{\mathcal P}^2 ) \mathcal I _{2\sigma} ^{22}] - 336 (1+\beta) f _{\mathcal P} m _Q \mathcal I _{3\mathcal P} ^{12}
		] \nneq + 56 (1+\beta)^2 m _{Q'}^2 m _Q (-1+\tilde \mu _{\mathcal P}^2) \mu _{\mathcal P} \mathcal I _{3\sigma}^{11}
	\} \\
	\Pi ^{(AA){\rm theo}} &= \frac{M^2}{1728\pi^2 m _{Q'}} \{
		9 M^2 (1-\beta)^2 m _Q \mu _{\mathcal P} \mathcal I _1^{12} \nneq + M^2 (-1+\beta) m _{Q'} [
			\mu _{\mathcal P} [9(-1+\beta) I_1^{12} + 56 (3+\beta) (-1+\tilde \mu _{\mathcal P}^2 ) \mathcal I _{2\sigma} ^{22}] 
			 - 336 (1+5\beta) f _{\mathcal P} m _Q \mathcal I _{3\mathcal P}^{12} 
		] \nneq + 56 (3+\beta) (1+3\beta) m _{Q'}^2 m _Q (-1+\tilde \mu _{\mathcal P}^2 ) \mu _{\mathcal P} I _{3\sigma}^{11}
 	\} \\
	\Pi ^{(SA){\rm theo}} &= \frac{M^2}{288\sqrt 6 \pi^2 m _{Q'}} \{
		9M^2 (-1+\beta) m _Q \mu _{\mathcal P} [ (-1+\beta) \mathcal I _1^{12} (\mathcal T(\alpha _i)) + 2 (3+\beta) \mathcal I _1^{12} (u \mathcal T (\alpha _i)) ] 
		\nneq + M^2 (-1+\beta) m _{Q'} [
			\mu _{\mathcal P} [ 9(-1+\beta) \mathcal I _{1}^{12} (\mathcal T (\alpha _i)) - 18 (1+3\beta) \mathcal I _1^{12} (u\mathcal T (\alpha _i)) - 56 (-1+\beta) (-1+\tilde \mu _{\mathcal P}^2) \mathcal I _{2\sigma}^{22} ] 
			\nneq -336(1+\beta) f _{\mathcal P} m _Q \mathcal I _{3\mathcal P}^{12} 
		] 
		+ 56 (1+\beta)^2 m _{Q'}^2 m _Q (-1+\tilde \mu _{\mathcal P}^2) \mu _{\mathcal P} \mathcal I _{3\sigma}^{11}
	\}
	\end{align}
where we have defined
\begin{align}
I _1 ^{ij} (f(u) \mathcal A (\alpha _i)) &= \int _{(m _Q + m _{Q'})^2} ^{s _0} \dif s\ \e{-s/M^2} \int  \dif \alpha\ \alpha ^{i-1} (1-\alpha)^{j-1} \delta\Big(
s - \Big(
\frac{m _Q^2}{\alpha} + \frac{m _{Q'}^2}{1-\alpha}
\Big)
\Big)
\nn \times \int _0 ^{u _0} \dif \alpha _1 \int _{1-u _0}^{1-\alpha _1} \dif \alpha _3\  \frac{f(\frac{u _0-\alpha _1}{\alpha _3})}{u _0 - \alpha _1} \frac{\dif \mathcal A (\alpha _1,1-\alpha _1 - \alpha _3,\alpha _3)}{\dif \alpha _3} \\
I _{2\sigma}^{ij} &= \varphi _\sigma (u _0) \int _{(m _Q + m _{Q'})^2} ^{s _0} \dif s\ \e{-s/M^2} \int  \dif \alpha\ \alpha ^{i-1} (1-\alpha)^{j-1} \delta\Big(
s - \Big(
\frac{m _Q^2}{\alpha} + \frac{m _{Q'}^2}{1-\alpha}
\Big)
\Big) \bb{-(i+j-1) - \frac{s}{M^2}} \\
I _{3\mathcal P}^{ij} &= \varphi _{\mathcal P}(u _0) \int _{(m _Q + m _{Q'})^2} ^{s _0} \dif s\ \e{-s/M^2} \int  \dif \alpha\ \alpha ^{i-1} (1-\alpha)^{j-1} \delta\Big(
s - \Big(
\frac{m _Q^2}{\alpha} + \frac{m _{Q'}^2}{1-\alpha}
\Big)
\Big)  \\
I _{3\sigma}^{ij} &= \varphi _{\sigma}(u _0) \int _{(m _Q + m _{Q'})^2} ^{s _0} \dif s\ \e{-s/M^2} \int  \dif \alpha\ \alpha ^{i-1} (1-\alpha)^{j-1} \delta\Big(
s - \Big(
\frac{m _Q^2}{\alpha} + \frac{m _{Q'}^2}{1-\alpha}
\Big)
\Big) 
\end{align}
\end{widetext}
where
\begin{align}
\alpha _{\rm max,min} &= \frac{1}{2s _0} [s _0 + m _Q ^2 - m _{Q'}^2 \pm ((s _0 + m _Q^2 - m _{Q'}^2)^2 \nneq - 4 s _0 m _Q^2)^{1/2}] \label{eq:alphaminmax}
\end{align}
and $ N $ is the normalization factor which is equal to $ 1/\sqrt 2 $ (1) for different (identical) heavy quark flavors. Here, we should note that we present the results for the pion case, where we take $ m _\pi^2 \to 0 $ for simplicity, retaining $ m _\pi^2 $ in the terms $ \frac{m _\pi^2}{m _u+m _d} $. For the kaon case, we take into account the contribution from all the terms.
\par 
As an example, we present the steps of calculations for one of the terms that appear in the calculation of the theoretical part of the correlation function, and the results for the remaining terms are presented in Appendix \hyperref[app:B]{B}.
\par 
We consider the term
\begin{align*}
\int \dif u \int \dif ^4 x \ \e{i(p+uq)x} \frac{K _i(m _Q\sqrt{-x^2})}{(\sqrt{-x^2})^i} \frac{K _j(m _{Q'}\sqrt{-x^2})}{(\sqrt{-x^2})^j} \varphi (u)
\end{align*}
where $ \varphi (u) $ is a generic two-particle DA of the pseudoscalar meson. Using the integral representation of the Bessel function, 
\begin{align}
\frac{K _i(m _Q\sqrt{-x^2})}{(\sqrt{-x^2})^i} = \frac 12 \int _0 ^\infty \dif t \ \frac{1}{t ^{i+1}} \e{-\frac{m _Q}{2} (t + x _E ^2/t)}
\end{align}
where $ x _E ^2 = - x ^2 $. Introduce the new variables $ a $ and $ b $ as $ a = \frac{2m _Q}{t} $ and $ b = \frac{2m _{Q'}}{t'} $. Performing the integration over $ \int \dif ^4 x _E $, we get
\begin{align}
&\frac 14 \frac{16}{(2m _{Q})^i} \frac{i}{(2m _{Q'})^j} \pi ^2 \int \dif u \int _0 ^\infty \dif a \int _0^\infty \dif b \ a ^{i-1} b ^{j-1} \varphi (u) \nneq\times \frac{1}{(a+b)^2} \e{- P^2/(a+b)} \e{-m _Q^2/a-m _{Q'}^2/b}
\end{align}
where $  P := p _E + q _E x $. Introducing the identity $ \int \dif \rho \ \delta (\rho-a-b) = 1 $, making a scale transformation $ a\to \rho \alpha $, $ b \to \rho \beta $, and performing the integration over $ \beta $, we obtain
\begin{align}
&\frac i 4 \frac{16\pi^2}{(2m _Q)^i (2m _{Q'})^j} \int _0^1 \dif u \int _0 ^\infty \dif \alpha \int _0^\infty \dif \rho \ \frac 1\rho \alpha ^{i-1} \nneq\times (1-\alpha)^{j-1} \rho ^{i+j-2} \e{-\frac{p^2 \bar u + (p+q)^2 u}{\rho}}  \e{-\frac{m _Q^2}{\rho\alpha} - \frac{m _{Q'}^2}{\rho(1-\alpha)}} \varphi (u)
\end{align}
Performing the Borel transformations over variables $ -p^2 $ and $ -(p+q)^2 $ with the help of the formula $ \hat B \e{-\alpha p^2} = \delta \pp{\frac{1}{M^2} - \alpha} $, we obtain
\begin{align}
&\frac i4 \frac{16\pi^2 (M^2)^{i+j}}{(2m _Q)^i(2m _{Q'})^j} \varphi (u _0) \int _0^\infty  \dif \alpha \ \alpha ^{i-1} (1-\alpha) ^{j-1}\nneq\times \e{-\pp{\frac{m _Q^2}{\alpha} + \frac{m _{Q'}^2}{1-\alpha}}/M^2}
\end{align}
Let $ s $ denote $ \frac{m _Q^2}{\alpha} + \frac{m _{Q'}^2}{1-\alpha} $. Equating this to $ s _0 $ in order to perform the subtraction of the continuum threshold, we can find the bounds of $ \alpha $. As a result, we get
\begin{align}
&\frac i4 \frac{16\pi^2 (M^2)^{i+j}}{(2m _Q)^i (2m _{Q'})^j} \varphi (u _0) \int_{(m_Q + m_{Q'})^2}^{s_0} \dif s \ \e{-s/M^2}\int \dif \alpha \nn\times \alpha ^{i-1} (1-\alpha) ^{j-1} \delta(s - \frac{m_Q^2}{\alpha} - \frac{m_{Q'}^2}{1-\alpha})
\end{align}
In these expressions, 
\begin{align}
M^2 = \frac{M _1^2 M _1^2}{M _1^2 + M _2^2}, \quad u _0 = \frac{M _1^2}{M _1^2 + M _2 ^2}
\end{align}
Since in our case the mass of the initial and final baryons are practically the same, we put $ M _1^2 = M _2^2 $, which gives us $ u _0 = 1/2 $. 
\par 
Matching the two representations of the correlation function for the relevant coupling constants, we obtain
\begin{align}
g _{B _1B _2\mathcal P} = -\frac{1}{\lambda _{B _1} \lambda _{B _2}} \e{(m _{B _1}^2 + m _{B _2}^2)/2M^2} \Pi ^{\rm theo}
\end{align}
\section{Numerical analysis}\label{sec:3}
In this section, we numerically analyze the LCSR for the strong coupling constants of the $ \pi $ and $ K $ mesons with the baryons $ \Xi _{cc} $, $ \Xi _{bb} $, $ \Xi _{bc} $, $ \Xi _{bc}' $, $ \Omega _{cc} $, $ \Omega _{bb} $, $ \Omega _{bc} $, and $ \Omega _{bc}' $ by using Package X \cite{ref:30}. The LCSR for the coupling constants $ g _{B _1 B _2 \mathcal P} ^{(SS)} $, $ g _{B _1B _2 \mathcal P} ^{(AA)} $, and $ g _{B _1 B _2 \mathcal P} ^{(SA)} $ include certain input parameters such as quark masses, the masses and decay constants of the pseudoscalar mesons $ \pi $ and $ K $, and the masses and residues of doubly heavy baryons. Some of these parameters are given in Table \ref{tab:1}.
\begin{table}
	\caption{Some of the input parameters used in our computations.\label{tab:1}}
	\begin{ruledtabular}
		\begin{tabular}
			{cccccc}
			Parameter & Value & Parameter & Value   \\
			\hline 
			$ m _s $ (1 GeV)			& 137 MeV 				& $ m _{\Xi _{cc}} $ 			& 3.72 GeV \cite{ref:15} \\ 
			$ m _c $ 					& 1.4 GeV 				& $ m _{\Xi _{bb}} $ 			& 9.96 GeV \cite{ref:15} \\
			$ m _b $ 					& 4.7 GeV 				& $ m _{\Xi _{bc}} $ 			& 6.72 GeV \cite{ref:15} \\
			$ m _\pi $ 					& 135 MeV 				& $ m _{\Xi' _{bc}} $ 			& 6.79 GeV \cite{ref:15} \\
			$ m _K $ 					& 495 MeV 				& $ m _{\Omega _{bb}} $ 		& 9.97 GeV \cite{ref:15} \\
										& 		 				& $ m _{\Omega _{cc}} $ 		& 3.73 GeV \cite{ref:15} \\
			$ f _{\pi} $				& 131 MeV 				& $ m _{\Omega _{bc}} $ 		& 6.75 GeV \cite{ref:15} \\
			$ f _K $					& 160 MeV 				& $ m _{\Omega' _{bc}} $		& 6.80 GeV \cite{ref:15} \\
			&&&\\
			$ \lambda _{\Xi _{cc}} $	& 0.16 \cite{ref:15}	& $ \lambda _{\Omega _{cc}} $	& 0.18 \cite{ref:15} \\
			$ \lambda _{\Xi _{bb}} $ 	& 0.44 \cite{ref:15} 	& $ \lambda _{\Omega _{bb}} $	& 0.45 \cite{ref:15}\\
			$ \lambda _{\Xi _{bc}} $	& 0.28 \cite{ref:15}	& $ \lambda _{\Omega _{bc}} $	& 0.29 \cite{ref:15}\\
			$ \lambda _{\Xi'_{bc}} $	& 0.30 \cite{ref:15} 	& $ \lambda _{\Omega'_{bc}} $	& 0.31 \cite{ref:15}
		\end{tabular}
	\end{ruledtabular}
\end{table}
Another set of essential input parameters are the pseudoscalar meson DAs of different twists. These DAs are given as follows:
\begin{align}
\varphi _{\mathcal P} (u) &= 6u\bar u \bb{1 + a _1 ^{\mathcal P} C _1(2u-1) + a _2 ^{\mathcal P} C _2^{3/2} (2u-1)} \\
\varphi _{P} (u) &= 1 + \pp{30 \eta _3 - \frac 52 \frac{1}{\mu _{\mathcal P}^2}} C _2^{1/2} (2u-1) \nneq + \pp{-3\eta _3 w _3 - \frac{27}{20} \frac{1}{\mu _{\mathcal P} ^2} - \frac{81}{10} \frac{1}{\mu _{\mathcal P}^2} a _2 ^{\mathcal P}} C _4^{1/2} (2u-1)\\
\varphi _\sigma (u) &= 6 u \bar u \Big[ 
	1 + \pp{
		5 \eta _3 - \frac 12 \eta _3 w _3 - \frac{7}{20} \mu _{\mathcal P}^2 - \frac 35 \mu _{\mathcal P}^2 a _2 ^{\mathcal P}	
	}\nneq\times C _2^{3/2} (2u-1)
\Big] \\
\mathcal T (\alpha _i) &= 360 \eta _3 \alpha _{\bar q} \alpha _q \alpha _g ^2 \bb{1 + w _3 \frac 12 (7 \alpha _g - 3)}
\end{align}
The $ C _n^k (x) $ are the Gegenbauer polynomials. The values of the parameters inside the distribution amplitudes at the renormalization scale of $ \mu = 1 {\rm\ GeV} $ are $ a _1^\pi = 0 $, $ a _2^\pi = 0.44 $, $ a _1 ^K = 0.06 $, $ a _2^K = 0.25 $, $ \eta _3 = 0.015 $, and $ w _3 = -3 $ for the pion and $ w _3 = -1.2 $ for the kaon. 
\par
\begin{figure}
	\includegraphics[width=.5\textwidth]{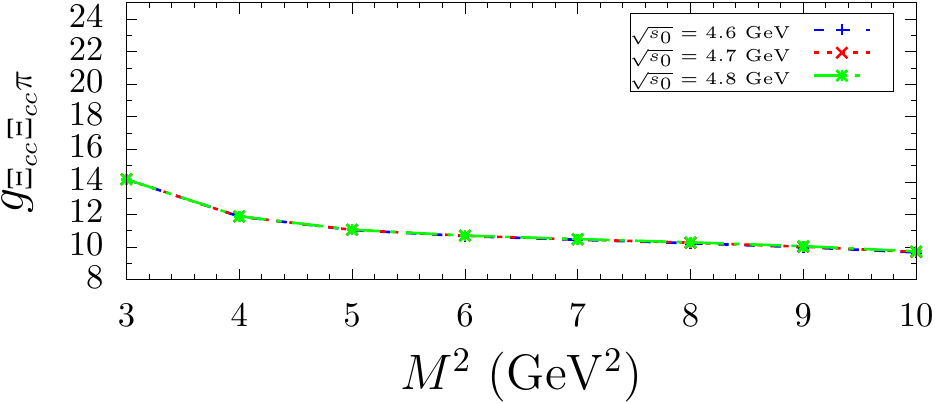}
	\caption{The dependence of the strong coupling constant $ g _{\Xi _{cc}\Xi _{cc}\pi} $ on $ M^2 $ at different $ s _0 $ values and $ \beta = 0.75 $.\label{fig:1}}
\end{figure}

\begin{figure}
	\includegraphics[width=.5\textwidth]{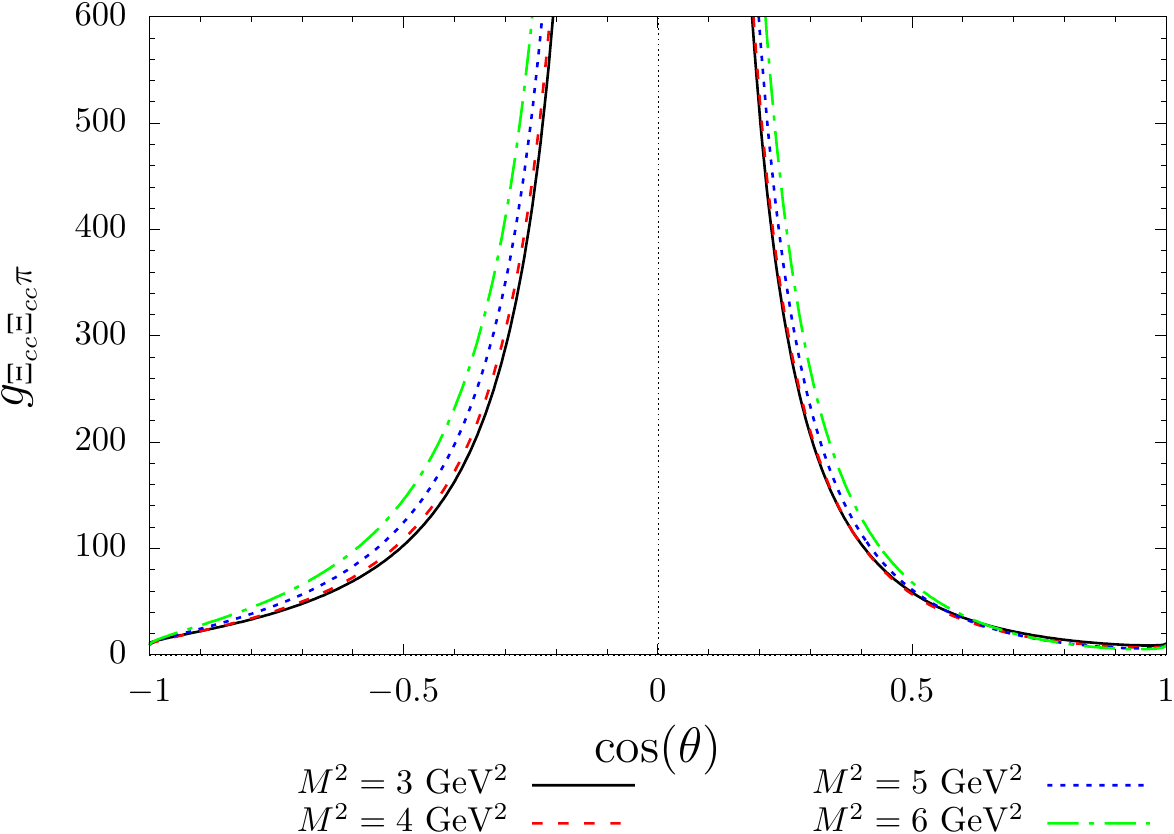}
	\caption{The dependence of the strong coupling constant $ g _{\Xi _{cc}\Xi _{cc}\pi} $ on $ \cos\theta $ at different $ M^2 $ values and $ \sqrt{s _0} = 4.6 {\rm\ GeV} $.\label{fig:2}}
\end{figure}

 
From sum rules for the coupling constant, we see that besides input parameters, they contain three auxiliary parameters: the Borel mass parameter, $ M^2 $, the continuum threshold, $ s _0 $, and the arbitrary parameter, $ \beta $, which appear in the expression for the interpolating current. Obviously, the measurable coupling constant should be independent of them. Therefore, we must find the working regions of these parameters for which the sum rules is reliable. The lowest bound of $ M^2 $ is obtained by requiring the highest-twist terms contributions should be reasonably small compared to the lowest-twist term contributions. The upper bound of $ M^2 $ is determined by demanding that the continuum contribution should be not too large. Consequently, we can find the working region of the Borel parameter $ M^2 $. The continuum threshold $ s _0 $ is obtained by requiring that the mass sum rules reproduce a 10\% accuracy of the mass of doubly heavy baryons. These conditions lead to the following values of $ M^2 $ and $ s _0 $ for its channel as follows: 
\begin{align}
	\begin{cases}
	\Xi _{cc} \Xi _{cc} \pi : \ M^2 \in [3,6] {\rm\ GeV^2}, \ \sqrt{s _0} = 4.6  {\rm\ GeV}\\
	\Xi _{bb} \Xi _{bb} \pi : \ M^2 \in [10,15] {\rm\ GeV^2}, \ \sqrt{s _0} =  10.9 {\rm\ GeV}\\
	\Omega _{bb} \Xi _{bb} K : \ M^2 \in [10,15] {\rm\ GeV^2}, \ \sqrt{s _0} =  10.9 {\rm\ GeV}\\
	\Omega _{cc} \Xi _{cc} K : \ M^2 \in [3,6] {\rm\ GeV^2}, \ \sqrt{s _0} =  4.6 {\rm\ GeV}\\
	\Omega _{bc} \Xi _{bc} K : \ M^2 \in [6,9] {\rm\ GeV^2}, \ \sqrt{s _0} = 7.5  {\rm\ GeV}\\
	\end{cases}
\end{align}
With the antisymmetric-antisymmetric current,
\begin{align}
\begin{cases}
\Xi' _{bc} \Xi' _{bc} \pi : \ M^2 \in [6,9] {\rm\ GeV^2}, \ \sqrt{s _0} =  7.5 {\rm\ GeV}\\
\Omega' _{bc} \Xi' _{bc} K :\ M^2 \in [6,9] {\rm\ GeV^2}, \ \sqrt{s _0} =  7.5 {\rm\ GeV}\\
\end{cases}
\end{align}
With the symmetric-antisymmetric current,
\begin{align}  
\begin{cases}
\Xi' _{bc} \Xi _{bc} \pi : \ M^2 \in [6,9] {\rm\ GeV^2}, \ \sqrt{s _0} =  7.5 {\rm\ GeV}\\
\Omega' _{bc} \Xi _{bc} K :\ M^2 \in [6,9] {\rm\ GeV^2}, \ \sqrt{s _0} =  7.5 {\rm\ GeV}
\end{cases}
\end{align} 
Our calculation shows that the twist-4 term contributions in these domains of $ M^2 $ at given values of $ s _0 $ does not exceed 14\% and higher states contribute at maximum 32\% for all considered channels. As an example, in Fig. \ref{fig:1}, we present the $ M^2 $ dependence of $ g _{\Xi _{cc} \Xi _{cc}\pi} $ at fixed values of $ s _0 $ and $ \beta $. 
Having the working regions of $ M^2 $ and $ s _0 $, we try to find the working region of $ \beta $. For this aim, we study the dependence of the strong coupling constant on $ \cos\theta $, where $ \beta = \tan \theta $. We will search for a domain for $ \beta $ such that the results are insensitive to the variation in $ \beta $. As an example, the dependence of the strong coupling constant $ g _{\Xi _{cc}\Xi _{cc}\pi} $ on $ \cos\theta $ at fixed values of $ M^2 $ and $ \sqrt{s _0} = 4.6 {\rm\ GeV} $ is presented in Fig. \ref{fig:2}. From Fig. \ref{fig:2}, one can see that when $ \cos\theta $ varies between 0.6 and 1, the coupling constant practically does not change and we deduce the values in Table \ref{tab:2}.
Performing similar calculations for the remaining of the strong coupling constants, we obtain the results that are summarized in Table \ref{tab:2}. The uncertainties are due to the variation of $ M^2 $, $ s _0 $, and errors in the values of the input parameters.
\par 
We would like to say a few words on the results obtained. In the case of symmetric currents, the difference in the values of the couplings constants of $ \Xi _{QQ} \Xi _{QQ} \pi $ and $ \Omega _{QQ} \Omega _{QQ} K $ are primarily due to the $ SU(3) $ symmetry violation. We mainly see that the $ SU(3) $ violation in the $ c $ sector is about 20\% but in the $ b $ sector, it is about 30\%--35\%. In the case of antisymmetric currents, the $ SU(3) $ violation is about 15\%--20\%; however, in the case of symmetric-antisymmetric current, the violation of $ SU(3) $ is about 35\%, similar to the antisymmetric-antisymmetric case.
\begin{table}
	\caption{The numerical values for the strong coupling constants.\label{tab:2}}
	\begin{ruledtabular}
		\begin{tabular}
			{ccc}
			& Channel & Strong coupling constant \\
			\hline 
			\multirow{5}{*}{$ SS $} & $ \Xi _{cc} \Xi _{cc} \pi $ & $ 10.03 \pm 0.52 $\\
			& $ \Xi _{bb} \Xi _{bb} \pi $ & $ 12.73 \pm 1.29 $\\
			& $ \Omega _{bb} \Xi _{bb} K $ & $ 17.40 \pm 1.89 $ \\
			& $ \Omega _{cc} \Xi _{cc} K $ & $ 12.50 \pm 0.75 $\\
			& $ \Omega _{bc} \Xi _{bc} K $ & $ 5.08 \pm 0.43 $\\
			\hline 
			\multirow{2}{*}{$ AA $}	& $ \Xi ' _{bc} \Xi ' _{bc} \pi $& $ 6.85 \pm 0.06 $\\
			& $ \Omega' _{bc} \Xi' _{bc} K $& $ 7.90 \pm 0.16 $\\
			\hline 
			\multirow{2}{*}{$ SA $}	& $ \Xi' _{bc} \Xi _{bc} \pi $& $ 1.49 \pm 0.10 $\\
			& $ \Omega' _{bc} \Xi _{bc} K $& $ 2.03 \pm 0.16 $ \\	
		\end{tabular}
	\end{ruledtabular}
\end{table}
Our final remark to this section is as follows. As we have already noted, the $ \Xi _{cc} \Xi _{cc} \pi $ and $ \Xi _{bb} \Xi _{bb} \pi $ coupling constants within the same framework is calculated in \cite{ref:29} and our results differ from the one given in \cite{ref:29}. 
In our opinion, these differences are due to the following
circumstances: (i) The main equation, Eq. (8) of \cite{ref:29} is
incorrect. It is due to the following simple fact. Let us consider the terms without $ t $ (in our case, it is $ \beta $) in Eq. (8). By using Eq. (11), from Eq. (8) with $ \Gamma = \gamma _5 $ or $ \Gamma = \gamma _\mu \gamma _5 $, immediately one gets that the terms which $ \phi _\pi $ and $ \phi _P $ do not contribute to the correlation function. But, these terms appear in Eq. (26), which seems highly strange. (ii) The continuum subtraction procedure
performed in \cite{ref:29} is inconsistent. This is due to the apparent fact that the variables $ z $ and $ s $ are related as $ \frac{m _1^2}{z} + \frac{m _2^2}{1-z} = s $. However, in [29], $ z $ and $ s $ are introduced as two independent variables, which is incorrect. Using these facts one
can conclude that the results of \cite{ref:29} are not reliable.

\begin{acknowledgements}
	One of the authors (H. I. A.) extends her appreciation to the Deanship of Scientific Research at Princess Nourah bint Abdulrahman University, where this research was funded by Grant No. 39-YR-1.
\end{acknowledgements}
\appendix 
\begin{widetext}
	\section{DISTRIBUTION AMPLITUDES OF THE PION AND KAON\label{app:A}}
	In this appendix, we present explicit expressions for the DAs of the $ \pi $ meson. For more information, see \cite{ref:31} and \cite{ref:32}. 
	\begin{align}
	\bracket{\mathcal P (p)}{\bar q(x) \gamma _\mu \gamma _5 q (0)}{0} &= -i f _{\mathcal P} p _\mu \int _0^1 \dif u \ \e{i\bar u px} \bb{\varphi _{\mathcal P} (u) + \frac{1}{16} m _{\mathcal P}^2 x^2 \hat A (u)} 
	- \frac i 2 f _{\mathcal P} m _{\mathcal P}^2 \frac{x _\mu}{px} \int _0^1 \dif u \ \e{i\bar u px} \hat B(u) 
	\end{align}
	\begin{align}
	\bracket{\mathcal P (p)}{\bar q (x) i \gamma _5 q (0)}{0} &= \mu _{\mathcal P} \int _0^1 \dif u \ \e{i\bar u px} \varphi _P(u) 
	\end{align}
	\begin{align}
	\bracket{\mathcal P (p)}{\bar q (x) \sigma _{\alpha\beta} \gamma _5 q (0)}{0} &= \frac i 6 \mu _{\mathcal P} (1-\tilde \mu _{\mathcal P}^2) (p _\alpha x _\beta - p _\beta x _\alpha) \int _0^1 \dif u \ \e{i\bar u px} \varphi _\sigma (u)
	\end{align}
	\begin{align}
	\bracket{\mathcal P (p)}{\bar q (x) \sigma _{\mu\nu} \gamma _5 g _s G _{\alpha\beta} (vx) q (0)}{0} &= i \mu _{\mathcal P} \Big\{
	p _\alpha p _\mu \bb{g _{\nu\beta} - \frac{1}{px} (p _\nu x _\beta + p _\beta x _\nu)} 
	 - p_\alpha p _\nu \bb{g _{\mu\beta} - \frac{1}{px} (p _\mu x _\beta + p _\beta x _\mu)} 
	\nnhp{= i \mu _{\mathcal P} \Big\{} - p _\beta p _\mu \bb{g _{\nu\alpha} - \frac{1}{px} (p _\nu x _\alpha + p _\alpha x _\nu)}
	 + p _\beta p _\nu \bb{g _{\mu\alpha} - \frac{1}{px} (p _\mu x _\alpha + p _\alpha x _\mu)}
	\Big\}\nneq \times \int \mathcal D \alpha\ \e{i(\alpha _{\bar q} + v \alpha _g)px} \mathcal T (\alpha _i)
	\end{align}
	\begin{align}
	\bracket{\mathcal P (p)}{\bar q (x) \gamma _\mu \gamma _5 g _s G _{\alpha\beta} (vx) q(0)}{0} &= p _\mu (p _\alpha x _\beta - p _\beta x _\alpha) \frac{1}{px} f _{\mathcal P} m _{\mathcal P}^2 \int \mathcal D \alpha\ \e{i(\alpha _{\bar q} + v \alpha _g)px} \mathcal A _\parallel (\alpha _i) \nneq 
	+\Big\{
	p _\beta \bb{g _{\mu\alpha} - \frac{1}{px} (p _\mu x _\alpha 
		+ p _\alpha x _\mu)}
	 - p _\alpha \bb{g _{\mu\beta} - \frac{1}{px} (p _\mu x _\beta + p _\beta x _\mu)}
	\Big\}  
	\nnhp{= +} \times  f _{\mathcal P} m _{\mathcal P} ^2 \int \mathcal D \alpha\ \e{i(\alpha _{\bar q} + v \alpha _g) px} \mathcal A _\perp (\alpha _i)
	\end{align}
	\begin{align}
	\bracket{\mathcal P (p)}{\bar q (x) \gamma _\mu i g _s G _{\alpha\beta} (vx) q(0)}{0} &= p _\mu (p _\alpha x _\beta - p _\beta x _\alpha) \frac{1}{px} f _{\mathcal P} m _{\mathcal P}^2 \int \mathcal D \alpha\ \e{i(\alpha _{\bar q} + v\alpha _g)px} \mathcal V _\parallel (\alpha _i)
	\nneq + \Big\{
	p _\beta \bb{g _{\mu\alpha} - \frac{1}{px} (p _\mu x _\alpha + p_\alpha x _\mu) } 
	 - p _\alpha \bb{g _{\mu\beta} - \frac{1}{px} (p _\mu x _\beta + p _\beta x _\mu)}
	\Big\} 
	\nnhp{= + }\times f _{\mathcal P} m _{\mathcal P} ^2 \int \mathcal D \alpha \ \e{i(\alpha _{\bar q} + v\alpha _g)px} \mathcal V _\perp (\alpha _i) 
	\end{align}
	where
	\begin{align}
	\mu _{\mathcal P} = f _{\mathcal P} \frac{m _{\mathcal P}^2}{m _{q _1} + m _{q _2}},\quad \tilde \mu _{\mathcal P} = \frac{m _{q _1} + m _{q _2}}{m _{\mathcal P}}
	\end{align}
	where $ m _{q _1} = m _u $ and $ m _{q _2} = m _d $ for the pion, and $ m _{q _1} = m _u $ and $ m _{q _2} = m _s $ for the kaon.
	Here, $ \varphi _{\mathcal P} (u) $, $ \hat A (u) $, $ \hat B (u) $, $ \varphi _P (u) $, $ \varphi _\sigma (u) $, $ \mathcal T (\alpha _i) $, $ \mathcal A _\perp (\alpha _i) $, $ \mathcal A _\parallel (\alpha _i) $, $ \mathcal V _\perp (\alpha _i) $, and $ \mathcal V _\parallel (\alpha _i) $ are the distribution amplitudes of the pseudoscalar meson with definite twist. 
	\section{THEORETICAL RESULTS IN COMPUTING THE CORRELATION FUNCTION\label{app:B}}
	In this appendix, we present the theoretical results that appear in the calculation of the correlation function from the QCD side. 
	\begin{align}
	&\int \dif u \int \dif ^4 x \ x _\mu \varphi (u) \frac{K _i(m _Q\sqrt{-x^2})}{(\sqrt{-x^2})^i} \frac{K _j(m _{Q'}\sqrt{-x^2})}{(\sqrt{-x^2})^j} \e { i (p+uq) x} \to \frac i4 \frac{16\pi^2 (M^2 )^{i+j-1} }{(2m _Q)^i (2m _{Q'})^j} \varphi (u _0) [-2i(p+u _0q) _\mu]
	\nn\hspace{6.6cm} \times \int _{(m_Q+m_{Q'})^2}^{s_0} \dif s\ e^{-s/M^2} \int \dif \alpha \ \alpha ^{i-1} (1-\alpha) ^{j-1} \delta(s-\frac{m_Q^2}{\alpha} - \frac{m_{Q'}^2}{1-\alpha})
	\end{align}
	\begin{align}
	& \int \dif u \int \dif ^4 x \ x^2 \varphi (u) \frac{K _i(m _{Q}\sqrt{-x^2})}{(\sqrt{-x^2})^i} \frac{K _j(m _{Q'}\sqrt{-x^2})}{(\sqrt{-x^2})^j} \e{i(p+uq)x} \to \frac 14 \frac{i}{(2m _Q)^i} \frac{16\pi^2}{(2m _{Q'})^j} 4 \varphi (u _0) (M^2)^{i+j-1}
	\nn \hspace{0.5cm}	\times \int_{(m_Q+m_{Q'})^2}^{s_0}\dif s\ \e{-s/M^2} \int \dif \alpha \ \alpha ^{i-1} (1-\alpha)^{j-1} \bb{- (i+j-1) - \pp{\frac{m _Q^2}{\alpha} + \frac{m _{Q'}^2}{1-\alpha}} /M^2}\delta(s - \frac{m_Q^2}{\alpha} - \frac{m_{Q'}^2}{1-\alpha})
	\end{align}
	\begin{align}
	&\frac{1}{qx} \int \dif u \int \dif ^4 x \ \frac{K _i(m _{Q}\sqrt{-x^2})}{(\sqrt{-x^2})^i} \frac{K _j(m _{Q'}\sqrt{-x^2})}{(\sqrt{-x^2})^j} \varphi (u) \e{i(p+uq)x} \to (-i) \frac i4 \frac{16\pi^2}{(2m _Q)^i (2m _{Q'})^j} (M^2)^{i+j} \Phi (u _0)
	\nn\hspace{6.5cm}\times  \int _{(m_Q+m_{Q'})^2}^{s_0} \dif s \ \e{-s/M^2} \int  \dif \alpha \ \alpha ^{i-1} (1-\alpha)^{j-1} \delta(s - \frac{m_Q^2}{\alpha} - \frac{m_{Q'}^2}{1-\alpha})
	\end{align}
	\begin{align}
	& \frac{x _\mu}{qx} \int \dif u \int \dif ^4 x \ \frac{K _i(m _{Q}\sqrt{-x^2})}{(\sqrt{-x^2})^i} \frac{K _j(m _{Q'}\sqrt{-x^2})}{(\sqrt{-x^2})^j} \varphi (u) \to \frac i4 \frac{16\pi^2}{(2m _Q)^i (2m _{Q'})^j} (-i) (M^2) ^{i+j-1} \Phi (u _0) [-2i (p+u _0 q) _\mu]
	\nn \hspace{6.5cm}\times \int_{(m_Q + m_{Q'})^2}^{s_0} \dif s \ \e{-s/M^2} \int \dif \alpha \ \alpha ^{i-1} (1-\alpha) ^{j-1} \delta (s - \frac{m_Q^2}{\alpha} - \frac{m_{Q'}^2}{1-\alpha})
	\end{align}
	\begin{align}
	& \int \dif u \int \dif ^4 x \int \dif x _1 \dif x _3 \ \e{i(p+q+ux _3)x} A (x _1,1-x _1-x _3,x _3) \frac{K _i(m _{Q}\sqrt{-x^2})}{(\sqrt{-x^2})^i} \frac{K _j(m _{Q'}\sqrt{-x^2})}{(\sqrt{-x^2})^j} \to \frac i4 \frac{16\pi^2 (M^2)^{i+j}}{(2m _Q)^i (2m _{Q'})^j}  
	\nn \times \int_{(m_Q + m_{Q'})^2} ^{s_0} \dif s \ \e{-s/M^2} \int \dif \alpha \int _0 ^{u _0} \dif x _1 \int _{u _0 - x _1}^{1-x _1} \dif x _3 \ \frac{1}{x _3} \delta(s - \frac{m_Q^2}{\alpha} - \frac{m_{Q'}^2}{1-\alpha}) 
	\alpha ^{i-1} (1-\alpha) ^{j-1} A (x _1,1-x _1-x _3,x _3) 
	\end{align}
	\begin{align}
	& \frac{1}{qx} \int \dif u \int \dif ^4 x \int \dif x _1 \dif x _3 \ \e{i(p+q+ux _3)x} A (x _1,1-x _1-x _3,x _3) \frac{K _i(m _{Q}\sqrt{-x^2})}{(\sqrt{-x^2})^i} \frac{K _j(m _{Q'}\sqrt{-x^2})}{(\sqrt{-x^2})^j}  \to (-i)\frac i4 \frac{16\pi^2 (M^2)^{i+j}}{(2m _Q)^i(2m _{Q'})^j}  
	\nn\hspace{1.3cm}\times \int_{(m_Q + m_{Q'})^2} ^{s_0} \dif s\ \e{-s/M^2} \int \dif \alpha \int _0^{u _0} \dif x _1 \int _{1-u _0}^{1-\alpha _1} \dif x _3 \ \frac{u _0-x _1}{x _3^2} 
	\delta (s - \frac{m_Q^2}{\alpha} - \frac{m_{Q'}^2}{1-\alpha})
	\hat A (x _1,1-x _1-x _3,x _3) 
	\end{align}
	where $ \hat A (x, 1-x _1-x _3,x _3) = \int _0^{x _3} \dif x \ A (x _1,1-x _1-x,x) $ and $ \Phi (u) = \int _0 ^u \dif v \ \phi (v) $.
\end{widetext}
\bibliography{refs}

\providecommand{\noopsort}[1]{}\providecommand{\singleletter}[1]{#1}%
\begin{thebibliography}{32}%
\makeatletter
\providecommand \@ifxundefined [1]{%
 \@ifx{#1\undefined}
}%
\providecommand \@ifnum [1]{%
 \ifnum #1\expandafter \@firstoftwo
 \else \expandafter \@secondoftwo
 \fi
}%
\providecommand \@ifx [1]{%
 \ifx #1\expandafter \@firstoftwo
 \else \expandafter \@secondoftwo
 \fi
}%
\providecommand \natexlab [1]{#1}%
\providecommand \enquote  [1]{``#1''}%
\providecommand \bibnamefont  [1]{#1}%
\providecommand \bibfnamefont [1]{#1}%
\providecommand \citenamefont [1]{#1}%
\providecommand \href@noop [0]{\@secondoftwo}%
\providecommand \href [0]{\begingroup \@sanitize@url \@href}%
\providecommand \@href[1]{\@@startlink{#1}\@@href}%
\providecommand \@@href[1]{\endgroup#1\@@endlink}%
\providecommand \@sanitize@url [0]{\catcode `\\12\catcode `\$12\catcode
  `\&12\catcode `\#12\catcode `\^12\catcode `\_12\catcode `\%12\relax}%
\providecommand \@@startlink[1]{}%
\providecommand \@@endlink[0]{}%
\providecommand \url  [0]{\begingroup\@sanitize@url \@url }%
\providecommand \@url [1]{\endgroup\@href {#1}{\urlprefix }}%
\providecommand \urlprefix  [0]{URL }%
\providecommand \Eprint [0]{\href }%
\providecommand \doibase [0]{https://doi.org/}%
\providecommand \selectlanguage [0]{\@gobble}%
\providecommand \bibinfo  [0]{\@secondoftwo}%
\providecommand \bibfield  [0]{\@secondoftwo}%
\providecommand \translation [1]{[#1]}%
\providecommand \BibitemOpen [0]{}%
\providecommand \bibitemStop [0]{}%
\providecommand \bibitemNoStop [0]{.\EOS\space}%
\providecommand \EOS [0]{\spacefactor3000\relax}%
\providecommand \BibitemShut  [1]{\csname bibitem#1\endcsname}%
\let\auto@bib@innerbib\@empty
\bibitem [{\citenamefont {Moinester}(1996)}]{ref:1}%
  \BibitemOpen
  \bibfield  {author} {\bibinfo {author} {\bibfnamefont {M.~A.}\ \bibnamefont
  {Moinester}},\ }\href {https://doi.org/10.1007/s002180050123} {\bibfield
  {journal} {\bibinfo  {journal} {Z. Phys. A}\ }\textbf {\bibinfo {volume}
  {355}},\ \bibinfo {pages} {349} (\bibinfo {year} {1996})}\BibitemShut
  {NoStop}%
\bibitem [{\citenamefont {Mattson}\ \emph {et~al.}(2002)\citenamefont {Mattson}
  \emph {et~al.}}]{ref:2}%
  \BibitemOpen
  \bibfield  {author} {\bibinfo {author} {\bibfnamefont {M.}~\bibnamefont
  {Mattson}} \emph {et~al.} (\bibinfo {collaboration} {SELEX Collaboration}),\
  }\href {https://doi.org/10.1103/physrevlett.89.112001} {\bibfield  {journal}
  {\bibinfo  {journal} {Phys. Rev. Lett.}\ }\textbf {\bibinfo {volume} {89}}
  (\bibinfo {year} {2002})}\BibitemShut {NoStop}%
\bibitem [{\citenamefont {Ocherashvili}\ \emph {et~al.}(2005)\citenamefont
  {Ocherashvili} \emph {et~al.}}]{ref:3}%
  \BibitemOpen
  \bibfield  {author} {\bibinfo {author} {\bibfnamefont {A.}~\bibnamefont
  {Ocherashvili}} \emph {et~al.} (\bibinfo {collaboration} {SELEX
  COllaboration}),\ }\href {https://doi.org/10.1016/j.physletb.2005.09.043}
  {\bibfield  {journal} {\bibinfo  {journal} {Phys. Lett. B}\ }\textbf
  {\bibinfo {volume} {628}},\ \bibinfo {pages} {18} (\bibinfo {year}
  {2005})}\BibitemShut {NoStop}%
\bibitem [{\citenamefont {Engelfried}\ \emph {et~al.}(2005)\citenamefont
  {Engelfried} \emph {et~al.}}]{ref:4}%
  \BibitemOpen
  \bibfield  {author} {\bibinfo {author} {\bibfnamefont {J.}~\bibnamefont
  {Engelfried}} \emph {et~al.} (\bibinfo {collaboration} {SELEX
  Collaboraton}),\ }\href {https://doi.org/10.1016/j.nuclphysa.2005.02.031}
  {\bibfield  {journal} {\bibinfo  {journal} {Nucl. Phys.}\ }\textbf {\bibinfo
  {volume} {A752}},\ \bibinfo {pages} {121} (\bibinfo {year}
  {2005})}\BibitemShut {NoStop}%
\bibitem [{ref(2017)}]{ref:5}%
  \BibitemOpen
  \href {https://doi.org/10.1103/physrevlett.119.112001} {\ \textbf {\bibinfo
  {volume} {119}},\ \bibinfo {pages} {112001} (\bibinfo {year}
  {2017})}\BibitemShut {NoStop}%
\bibitem [{\citenamefont {Aaij}\ \emph {et~al.}(2018)\citenamefont {Aaij} \emph
  {et~al.}}]{ref:6}%
  \BibitemOpen
  \bibfield  {author} {\bibinfo {author} {\bibfnamefont {R.}~\bibnamefont
  {Aaij}} \emph {et~al.} (\bibinfo {collaboration} {LHCb COllaboration}),\
  }\href {https://doi.org/10.1103/physrevlett.121.162002} {\bibfield  {journal}
  {\bibinfo  {journal} {Phys. Rev. Lett.}\ }\textbf {\bibinfo {volume} {121}},\
  \bibinfo {pages} {162002} (\bibinfo {year} {2018})}\BibitemShut {NoStop}%
\bibitem [{\citenamefont {Cerri}\ \emph {et~al.}()\citenamefont {Cerri} \emph
  {et~al.}}]{ref:7}%
  \BibitemOpen
  \bibfield  {author} {\bibinfo {author} {\bibfnamefont {A.}~\bibnamefont
  {Cerri}} \emph {et~al.},\ }\href@noop {} {}\Eprint
  {https://arxiv.org/abs/1812.07638} {arXiv:1812.07638} \BibitemShut {NoStop}%
\bibitem [{\citenamefont {Aaij}\ \emph {et~al.}(2019)\citenamefont {Aaij} \emph
  {et~al.}}]{ref:8}%
  \BibitemOpen
  \bibfield  {author} {\bibinfo {author} {\bibfnamefont {R.}~\bibnamefont
  {Aaij}} \emph {et~al.} (\bibinfo {collaboration} {LHCb Collaboration}),\
  }\href {https://doi.org/10.1007/s11433-019-1471-8} {\bibfield  {journal}
  {\bibinfo  {journal} {Sci. China Phys. Mech. Astron.}\ }\textbf {\bibinfo
  {volume} {63}} (\bibinfo {year} {2019})}\BibitemShut {NoStop}%
\bibitem [{\citenamefont {Yoshida}\ \emph {et~al.}(2015)\citenamefont
  {Yoshida}, \citenamefont {Hiyama}, \citenamefont {Hosaka}, \citenamefont
  {Oka},\ and\ \citenamefont {Sadato}}]{ref:9}%
  \BibitemOpen
  \bibfield  {author} {\bibinfo {author} {\bibfnamefont {T.}~\bibnamefont
  {Yoshida}}, \bibinfo {author} {\bibfnamefont {E.}~\bibnamefont {Hiyama}},
  \bibinfo {author} {\bibfnamefont {A.}~\bibnamefont {Hosaka}}, \bibinfo
  {author} {\bibfnamefont {M.}~\bibnamefont {Oka}},\ and\ \bibinfo {author}
  {\bibfnamefont {K.}~\bibnamefont {Sadato}},\ }\href
  {https://doi.org/10.1103/physrevd.92.114029} {\bibfield  {journal} {\bibinfo
  {journal} {Phys. Rev. D}\ }\textbf {\bibinfo {volume} {92}},\ \bibinfo
  {pages} {114029} (\bibinfo {year} {2015})}\BibitemShut {NoStop}%
\bibitem [{\citenamefont {Shah}\ \emph {et~al.}(2016)\citenamefont {Shah},
  \citenamefont {Thakkar},\ and\ \citenamefont {Rai}}]{ref:10}%
  \BibitemOpen
  \bibfield  {author} {\bibinfo {author} {\bibfnamefont {Z.}~\bibnamefont
  {Shah}}, \bibinfo {author} {\bibfnamefont {K.}~\bibnamefont {Thakkar}},\ and\
  \bibinfo {author} {\bibfnamefont {A.~K.}\ \bibnamefont {Rai}},\ }\href
  {https://doi.org/10.1140/epjc/s10052-016-4379-z} {\bibfield  {journal}
  {\bibinfo  {journal} {Eur. Phys. J. C}\ }\textbf {\bibinfo {volume} {76}},\
  \bibinfo {pages} {530} (\bibinfo {year} {2016})}\BibitemShut {NoStop}%
\bibitem [{\citenamefont {Brown}\ \emph {et~al.}(2014)\citenamefont {Brown},
  \citenamefont {Detmold}, \citenamefont {Meinel},\ and\ \citenamefont
  {Orginos}}]{ref:11}%
  \BibitemOpen
  \bibfield  {author} {\bibinfo {author} {\bibfnamefont {Z.~S.}\ \bibnamefont
  {Brown}}, \bibinfo {author} {\bibfnamefont {W.}~\bibnamefont {Detmold}},
  \bibinfo {author} {\bibfnamefont {S.}~\bibnamefont {Meinel}},\ and\ \bibinfo
  {author} {\bibfnamefont {K.}~\bibnamefont {Orginos}},\ }\href
  {https://doi.org/10.1103/physrevd.90.094507} {\bibfield  {journal} {\bibinfo
  {journal} {Phys. Rev. D}\ }\textbf {\bibinfo {volume} {90}},\ \bibinfo
  {pages} {094507} (\bibinfo {year} {2014})}\BibitemShut {NoStop}%
\bibitem [{\citenamefont {P{\'{e}}rez-Rubio}\ \emph {et~al.}(2015)\citenamefont
  {P{\'{e}}rez-Rubio}, \citenamefont {Collins},\ and\ \citenamefont
  {Bali}}]{ref:12}%
  \BibitemOpen
  \bibfield  {author} {\bibinfo {author} {\bibfnamefont {P.}~\bibnamefont
  {P{\'{e}}rez-Rubio}}, \bibinfo {author} {\bibfnamefont {S.}~\bibnamefont
  {Collins}},\ and\ \bibinfo {author} {\bibfnamefont {G.~S.}\ \bibnamefont
  {Bali}},\ }\href {https://doi.org/10.1103/physrevd.92.034504} {\bibfield
  {journal} {\bibinfo  {journal} {Phys. Rev. D}\ }\textbf {\bibinfo {volume}
  {92}},\ \bibinfo {pages} {034504} (\bibinfo {year} {2015})}\BibitemShut
  {NoStop}%
\bibitem [{\citenamefont {Zhang}\ and\ \citenamefont {Huang}(2008)}]{ref:13}%
  \BibitemOpen
  \bibfield  {author} {\bibinfo {author} {\bibfnamefont {J.-R.}\ \bibnamefont
  {Zhang}}\ and\ \bibinfo {author} {\bibfnamefont {M.-Q.}\ \bibnamefont
  {Huang}},\ }\href {https://doi.org/10.1103/physrevd.78.094007} {\bibfield
  {journal} {\bibinfo  {journal} {Phys. Rev. D}\ }\textbf {\bibinfo {volume}
  {78}},\ \bibinfo {pages} {094007} (\bibinfo {year} {2008})}\BibitemShut
  {NoStop}%
\bibitem [{\citenamefont {Wang}(2010{\natexlab{a}})}]{ref:14}%
  \BibitemOpen
  \bibfield  {author} {\bibinfo {author} {\bibfnamefont {Z.-G.}\ \bibnamefont
  {Wang}},\ }\href {https://doi.org/10.1140/epja/i2010-11004-3} {\bibfield
  {journal} {\bibinfo  {journal} {Eur. Phys. J. A}\ }\textbf {\bibinfo {volume}
  {45}},\ \bibinfo {pages} {267} (\bibinfo {year}
  {2010}{\natexlab{a}})}\BibitemShut {NoStop}%
\bibitem [{\citenamefont {Aliev}\ \emph {et~al.}(2012)\citenamefont {Aliev},
  \citenamefont {Azizi},\ and\ \citenamefont {Savc{\i}}}]{ref:15}%
  \BibitemOpen
  \bibfield  {author} {\bibinfo {author} {\bibfnamefont {T.}~\bibnamefont
  {Aliev}}, \bibinfo {author} {\bibfnamefont {K.}~\bibnamefont {Azizi}},\ and\
  \bibinfo {author} {\bibfnamefont {M.}~\bibnamefont {Savc{\i}}},\ }\href
  {https://doi.org/10.1016/j.nuclphysa.2012.09.009} {\bibfield  {journal}
  {\bibinfo  {journal} {Nucl. Phys.}\ }\textbf {\bibinfo {volume} {A895}},\
  \bibinfo {pages} {59} (\bibinfo {year} {2012})}\BibitemShut {NoStop}%
\bibitem [{\citenamefont {Aliev}\ \emph {et~al.}(2013)\citenamefont {Aliev},
  \citenamefont {Azizi},\ and\ \citenamefont {Savci}}]{ref:16}%
  \BibitemOpen
  \bibfield  {author} {\bibinfo {author} {\bibfnamefont {T.~M.}\ \bibnamefont
  {Aliev}}, \bibinfo {author} {\bibfnamefont {K.}~\bibnamefont {Azizi}},\ and\
  \bibinfo {author} {\bibfnamefont {M.}~\bibnamefont {Savci}},\ }\href
  {https://doi.org/10.1088/0954-3899/40/6/065003} {\bibfield  {journal}
  {\bibinfo  {journal} {J. Phys. G}\ }\textbf {\bibinfo {volume} {40}},\
  \bibinfo {pages} {065003} (\bibinfo {year} {2013})}\BibitemShut {NoStop}%
\bibitem [{\citenamefont {Albuquerque}\ and\ \citenamefont
  {Narison}(2010)}]{ref:17}%
  \BibitemOpen
  \bibfield  {author} {\bibinfo {author} {\bibfnamefont {R.}~\bibnamefont
  {Albuquerque}}\ and\ \bibinfo {author} {\bibfnamefont {S.}~\bibnamefont
  {Narison}},\ }\href {https://doi.org/10.1016/j.nuclphysbps.2010.10.068}
  {\bibfield  {journal} {\bibinfo  {journal} {Nucl. Phys. B Proc. Suppl.}\
  }\textbf {\bibinfo {volume} {207-208}},\ \bibinfo {pages} {265} (\bibinfo
  {year} {2010})}\BibitemShut {NoStop}%
\bibitem [{\citenamefont {Wang}(2010{\natexlab{b}})}]{ref:18}%
  \BibitemOpen
  \bibfield  {author} {\bibinfo {author} {\bibfnamefont {Z.-G.}\ \bibnamefont
  {Wang}},\ }\href {https://doi.org/10.1140/epjc/s10052-010-1357-8} {\bibfield
  {journal} {\bibinfo  {journal} {Eur. Phys. J. C}\ }\textbf {\bibinfo {volume}
  {68}},\ \bibinfo {pages} {459} (\bibinfo {year}
  {2010}{\natexlab{b}})}\BibitemShut {NoStop}%
\bibitem [{\citenamefont {Yu}\ and\ \citenamefont {Guo}(2019)}]{ref:19}%
  \BibitemOpen
  \bibfield  {author} {\bibinfo {author} {\bibfnamefont {Q.-X.}\ \bibnamefont
  {Yu}}\ and\ \bibinfo {author} {\bibfnamefont {X.-H.}\ \bibnamefont {Guo}},\
  }\href {https://doi.org/10.1016/j.nuclphysb.2019.114727} {\bibfield
  {journal} {\bibinfo  {journal} {Nucl. Phys.}\ }\textbf {\bibinfo {volume}
  {B947}},\ \bibinfo {pages} {114727} (\bibinfo {year} {2019})}\BibitemShut
  {NoStop}%
\bibitem [{\citenamefont {Weng}\ \emph {et~al.}(2018)\citenamefont {Weng},
  \citenamefont {Chen},\ and\ \citenamefont {Deng}}]{ref:20}%
  \BibitemOpen
  \bibfield  {author} {\bibinfo {author} {\bibfnamefont {X.-Z.}\ \bibnamefont
  {Weng}}, \bibinfo {author} {\bibfnamefont {X.-L.}\ \bibnamefont {Chen}},\
  and\ \bibinfo {author} {\bibfnamefont {W.-Z.}\ \bibnamefont {Deng}},\ }\href
  {https://doi.org/10.1103/physrevd.97.054008} {\bibfield  {journal} {\bibinfo
  {journal} {Phys. Rev. D}\ }\textbf {\bibinfo {volume} {97}},\ \bibinfo
  {pages} {054008} (\bibinfo {year} {2018})}\BibitemShut {NoStop}%
\bibitem [{\citenamefont {Shi}\ \emph {et~al.}(2020)\citenamefont {Shi},
  \citenamefont {Wang},\ and\ \citenamefont {Zhao}}]{ref:21}%
  \BibitemOpen
  \bibfield  {author} {\bibinfo {author} {\bibfnamefont {Y.-J.}\ \bibnamefont
  {Shi}}, \bibinfo {author} {\bibfnamefont {W.}~\bibnamefont {Wang}},\ and\
  \bibinfo {author} {\bibfnamefont {Z.-X.}\ \bibnamefont {Zhao}},\ }\href
  {https://doi.org/10.1140/epjc/s10052-020-8096-2} {\bibfield  {journal}
  {\bibinfo  {journal} {Eur. Phys. J. C}\ }\textbf {\bibinfo {volume} {80}},\
  \bibinfo {pages} {568} (\bibinfo {year} {2020})}\BibitemShut {NoStop}%
\bibitem [{\citenamefont {Shi}\ \emph {et~al.}(2019)\citenamefont {Shi},
  \citenamefont {Xing},\ and\ \citenamefont {Zhao}}]{ref:22}%
  \BibitemOpen
  \bibfield  {author} {\bibinfo {author} {\bibfnamefont {Y.-J.}\ \bibnamefont
  {Shi}}, \bibinfo {author} {\bibfnamefont {Y.}~\bibnamefont {Xing}},\ and\
  \bibinfo {author} {\bibfnamefont {Z.-X.}\ \bibnamefont {Zhao}},\ }\href
  {https://doi.org/10.1140/epjc/s10052-019-7014-y} {\bibfield  {journal}
  {\bibinfo  {journal} {Eur. Phys. J. C}\ }\textbf {\bibinfo {volume} {79}},\
  \bibinfo {pages} {501} (\bibinfo {year} {2019})}\BibitemShut {NoStop}%
\bibitem [{\citenamefont {Hu}\ and\ \citenamefont {Shi}(2020)}]{ref:23}%
  \BibitemOpen
  \bibfield  {author} {\bibinfo {author} {\bibfnamefont {X.-H.}\ \bibnamefont
  {Hu}}\ and\ \bibinfo {author} {\bibfnamefont {Y.-J.}\ \bibnamefont {Shi}},\
  }\href {https://doi.org/10.1140/epjc/s10052-020-7635-1} {\bibfield  {journal}
  {\bibinfo  {journal} {Eur. Phys. J. C}\ }\textbf {\bibinfo {volume} {80}},\
  \bibinfo {pages} {56} (\bibinfo {year} {2020})}\BibitemShut {NoStop}%
\bibitem [{\citenamefont {Wang}\ \emph {et~al.}(2017)\citenamefont {Wang},
  \citenamefont {Yu},\ and\ \citenamefont {Zhao}}]{ref:24}%
  \BibitemOpen
  \bibfield  {author} {\bibinfo {author} {\bibfnamefont {W.}~\bibnamefont
  {Wang}}, \bibinfo {author} {\bibfnamefont {F.-S.}\ \bibnamefont {Yu}},\ and\
  \bibinfo {author} {\bibfnamefont {Z.-X.}\ \bibnamefont {Zhao}},\ }\href
  {https://doi.org/10.1140/epjc/s10052-017-5360-1} {\bibfield  {journal}
  {\bibinfo  {journal} {Eur. Phys. J. C}\ }\textbf {\bibinfo {volume} {77}},\
  \bibinfo {pages} {781} (\bibinfo {year} {2017})}\BibitemShut {NoStop}%
\bibitem [{\citenamefont {Albertus}\ \emph {et~al.}(2007)\citenamefont
  {Albertus}, \citenamefont {Hern{\'{a}}ndez}, \citenamefont {Nieves},\ and\
  \citenamefont {Verde-Velasco}}]{ref:25}%
  \BibitemOpen
  \bibfield  {author} {\bibinfo {author} {\bibfnamefont {C.}~\bibnamefont
  {Albertus}}, \bibinfo {author} {\bibfnamefont {E.}~\bibnamefont
  {Hern{\'{a}}ndez}}, \bibinfo {author} {\bibfnamefont {J.}~\bibnamefont
  {Nieves}},\ and\ \bibinfo {author} {\bibfnamefont {J.~M.}\ \bibnamefont
  {Verde-Velasco}},\ }\href {https://doi.org/10.1140/epja/i2007-10364-y}
  {\bibfield  {journal} {\bibinfo  {journal} {Eur. Phys. J. A}\ }\textbf
  {\bibinfo {volume} {32}},\ \bibinfo {pages} {183} (\bibinfo {year}
  {2007})}\BibitemShut {NoStop}%
\bibitem [{\citenamefont {Ebert}\ \emph {et~al.}(2004)\citenamefont {Ebert},
  \citenamefont {Faustov}, \citenamefont {Galkin},\ and\ \citenamefont
  {Martynenko}}]{ref:26}%
  \BibitemOpen
  \bibfield  {author} {\bibinfo {author} {\bibfnamefont {D.}~\bibnamefont
  {Ebert}}, \bibinfo {author} {\bibfnamefont {R.~N.}\ \bibnamefont {Faustov}},
  \bibinfo {author} {\bibfnamefont {V.~O.}\ \bibnamefont {Galkin}},\ and\
  \bibinfo {author} {\bibfnamefont {A.~P.}\ \bibnamefont {Martynenko}},\ }\href
  {https://doi.org/10.1103/physrevd.70.014018} {\bibfield  {journal} {\bibinfo
  {journal} {Phys. Rev. D}\ }\textbf {\bibinfo {volume} {70}},\ \bibinfo
  {pages} {014018} (\bibinfo {year} {2004})}\BibitemShut {NoStop}%
\bibitem [{\citenamefont {Gutsche}\ \emph {et~al.}(2019)\citenamefont
  {Gutsche}, \citenamefont {Ivanov}, \citenamefont {K\"{o}rner}, \citenamefont
  {Lyubovitskij},\ and\ \citenamefont {Tyulemissov}}]{ref:27}%
  \BibitemOpen
  \bibfield  {author} {\bibinfo {author} {\bibfnamefont {T.}~\bibnamefont
  {Gutsche}}, \bibinfo {author} {\bibfnamefont {M.~A.}\ \bibnamefont {Ivanov}},
  \bibinfo {author} {\bibfnamefont {J.~G.}\ \bibnamefont {K\"{o}rner}},
  \bibinfo {author} {\bibfnamefont {V.~E.}\ \bibnamefont {Lyubovitskij}},\ and\
  \bibinfo {author} {\bibfnamefont {Z.}~\bibnamefont {Tyulemissov}},\ }\href
  {https://doi.org/10.1103/physrevd.100.114037} {\bibfield  {journal} {\bibinfo
   {journal} {Phys. Rev. D}\ }\textbf {\bibinfo {volume} {100}},\ \bibinfo
  {pages} {114037} (\bibinfo {year} {2019})}\BibitemShut {NoStop}%
\bibitem [{\citenamefont {Balitsky}\ \emph {et~al.}(1989)\citenamefont
  {Balitsky}, \citenamefont {Braun},\ and\ \citenamefont
  {Kolesnichenko}}]{ref:28}%
  \BibitemOpen
  \bibfield  {author} {\bibinfo {author} {\bibfnamefont {I.}~\bibnamefont
  {Balitsky}}, \bibinfo {author} {\bibfnamefont {V.}~\bibnamefont {Braun}},\
  and\ \bibinfo {author} {\bibfnamefont {A.}~\bibnamefont {Kolesnichenko}},\
  }\href {https://doi.org/10.1016/0550-3213(89)90570-1} {\bibfield  {journal}
  {\bibinfo  {journal} {Nucl. Phys.}\ }\textbf {\bibinfo {volume} {B312}},\
  \bibinfo {pages} {509} (\bibinfo {year} {1989})}\BibitemShut {NoStop}%
\bibitem [{\citenamefont {Olamaei}\ \emph {et~al.}(2020)\citenamefont
  {Olamaei}, \citenamefont {Azizi},\ and\ \citenamefont {Rostami}}]{ref:29}%
  \BibitemOpen
  \bibfield  {author} {\bibinfo {author} {\bibfnamefont {A.~R.}\ \bibnamefont
  {Olamaei}}, \bibinfo {author} {\bibfnamefont {K.}~\bibnamefont {Azizi}},\
  and\ \bibinfo {author} {\bibfnamefont {S.}~\bibnamefont {Rostami}},\ }\href
  {https://doi.org/10.1140/epjc/s10052-020-8194-1} {\bibfield  {journal}
  {\bibinfo  {journal} {Eur. Phys. J. C}\ }\textbf {\bibinfo {volume} {80}},\
  \bibinfo {pages} {613} (\bibinfo {year} {2020})}\BibitemShut {NoStop}%
\bibitem [{\citenamefont {Patel}(2015)}]{ref:30}%
  \BibitemOpen
  \bibfield  {author} {\bibinfo {author} {\bibfnamefont {H.~H.}\ \bibnamefont
  {Patel}},\ }\href {https://doi.org/10.1016/j.cpc.2015.08.017} {\bibfield
  {journal} {\bibinfo  {journal} {Comput. Phys. Commun.}\ }\textbf {\bibinfo
  {volume} {197}},\ \bibinfo {pages} {276} (\bibinfo {year}
  {2015})}\BibitemShut {NoStop}%
\bibitem [{ref()}]{ref:31}%
  \BibitemOpen
  \href@noop {} {\ }\bibinfo {note} {\hspace{-1mm}P. Ball,
  \href{https://doi.org/10.1088/1126-6708/1999/01/010}{J. High Energy Phys. 01
  (1999) 010}.}\BibitemShut {Stop}%
\bibitem [{\citenamefont {Ball}\ and\ \citenamefont {Zwicky}(2005)}]{ref:32}%
  \BibitemOpen
  \bibfield  {author} {\bibinfo {author} {\bibfnamefont {P.}~\bibnamefont
  {Ball}}\ and\ \bibinfo {author} {\bibfnamefont {R.}~\bibnamefont {Zwicky}},\
  }\href {https://doi.org/10.1103/physrevd.71.014015} {\bibfield  {journal}
  {\bibinfo  {journal} {Phys. Rev. D}\ }\textbf {\bibinfo {volume} {71}},\
  \bibinfo {pages} {014015} (\bibinfo {year} {2005})}\BibitemShut {NoStop}%
\end{thebibliography}%

\end{document}